\documentclass[10pt,journal,compsoc]{IEEEtran}
\usepackage{amsmath}
\usepackage{graphicx}
\usepackage{fixltx2e}
\usepackage{tabularx}
\usepackage[ruled,noline]{algorithm2e}

\usepackage{graphicx}

\SetAlFnt{\footnotesize}

\newcommand{\ie}{\unskip, i.\,e.,\ }
\SetKwProg{Fn}{Function}{}{}
\SetKwComment{Comment}{$\triangleright$\ }{}
\SetKwInOut{Parameter}{parameter}

\ifCLASSOPTIONcompsoc
  \usepackage[nocompress]{cite}
\else
  \usepackage{cite}
\fi

\ifCLASSINFOpdf

\else

\fi

\begin{document}
\bstctlcite{IEEEexample:BSTcontrol}

\title{Forecasting Cyber Attacks with Imbalanced Data Sets and Different Time Granularities}

\author{
    \IEEEauthorblockN{Ahmet Okutan\IEEEauthorrefmark{1}, Shanchieh Jay Yang\IEEEauthorrefmark{1}, Katie McConky\IEEEauthorrefmark{2}}
    
    \IEEEauthorblockA{\IEEEauthorrefmark{1}Computer Engineering, Rochester Institute of Technology, Rochester, NY, USA.}
    \\
    \IEEEauthorblockA{\IEEEauthorrefmark{2}Industrial and Systems Engineering, Rochester Institute of Technology, Rochester, NY, USA.}
}

\IEEEtitleabstractindextext{%
\begin{abstract}
If cyber incidents are predicted a reasonable amount of time before they occur, defensive actions to prevent their destructive effects could be planned. Unfortunately, most of the time we do not have enough observables of the malicious activities before they are already under way. Therefore, this work suggests to use unconventional signals extracted from various data sources with different time granularities to predict cyber incidents for target entities. A Bayesian network is used to predict cyber attacks where the unconventional signals are used as indicative random variables. This work also develops a novel minority class over sampling technique to improve cyber attack prediction on imbalanced data sets. The results show that depending on the selected time granularity, the unconventional signals are able to predict cyber attacks for the anonimyzed target organization even though the signals are not explicitly related to that organization. Furthermore, the minority over sampling approach developed achieves better performance compared to the existing filtering techniques in the literature.  
    
\end{abstract}

}

\maketitle

\section{Introduction}

The number and diversity of cyber attacks have increased recently compared to the previous years. Denial of Service (DOS), Malware, attack on Internet facing service (Defacement), malicious Email, and malicious URL are some of the attack types that organizations and Internet users commonly observe. PwC reports that the number of security incidents increased by 38\% in 2015 across all industries \cite{pwc}. Prediction of cyber attacks before they are executed is very important to take necessary defensive actions beforehand. This work proposes to forecast cyber attacks towards an entity by using unconventional signals from various data sources that may or may not be related to that target entity. For unconventional signals, this paper uses social media \ie Twitter and the open source GDELT \cite{gdelt} project that follows the media from almost every corner of the world, in over 100 languages, and almost every moment of each day.

The idea of using unconventional signals to forecast cyber events is based on the premise that the signals are not directly linked to specific exploits or vulnerabilities within the targeted organization. There might be many potentially useful unconventional signals, but each signal may not be particularly effective by itself. Furthermore, continuous enhancement to condition or aggregate these signals is not trivial and requires a significant amount of research. This work proposes a methodology based on Bayesian networks that can treat a variety of unconventional signals to forecast events that do not necessarily have balanced positive and negative ground truth instances. The feasibility of the proposed approach is demonstrated using the selected unconventional signals. Additional and potentially better signals may be used in conjunction with the method presented in this paper.

The main motivation behind this work is to address three important
cyber security problems: Use unconventional signals rather than the direct cyber observables to forecast cyber attacks, research the significance of changing time granularities for signals, and improve the performance of cyber attack forecasting on imbalanced cyber data sets. First, unlike some cyber attack prediction techniques which analyze the results of cyber incidents or suggest approaches to react to the attacks already underway, this research focuses on unconventional signals that have a potential to be indicative of a cyber incident towards a target entity. Bakdash \textit{et al.} \cite{Bakdash}  emphasize the importance of moving from reactive and passive defense systems to more proactive ones and show that the frequency of attacks from the previous week can be used to predict the number of attacks for the next week. Ramakrishnan \textit{et al.} \cite{Ramakrishnan:2014} develop an automated and continuously running system named EMBERS (Early Model Based Event Recognition using Surrogates) to forecast civil unrest events. They use open source indicators such as tweets, news sources, blogs, economic indicators, and other data sources as evidence and show that EMBERS is able to successfully forecast significant public unrest events across 10 countries of Latin America.

While using unconventional signals to forecast future cyber incidents, one needs to decide on the aggregation period of each signal and the entity ground truth, in order to generate training and test sets for target entities. Assuming $t$ represents a time stamp and $t_x$ and $t_g$ represent time granularities for the unconventional signals and the entity ground truth respectively, we investigate if signals aggregated over a period of time between $t - t_x$ and $t$ could be indicative of cyber events for the succeding time period starting at $t$ and ending at $t + t_g$. For instance, if $t_x$ is one week and $t_g$ is 24 hours, signals averaged over the last week could be used to predict attacks that will occur in the next 24 hours. Our previous study \cite{Okutan2017} calculated unconventional signals on a daily basis and used them with the ground truth data of the following day to predict cyber attacks before they are observed. This paper uses different time granularities for the unconventional signals ($t_x$) and the entity ground truth ($t_g$) and shows that the performance of a prediction model is dependent on the selected $t_x$ and $t_g$. 

Depending on the update frequency of a particular data source, choosing a very small or large $t_x$ affects the indicative power or quality of a specific signal. Similarly, contingent upon the type or reputation of an entity, the number and frequency of cyber incidents observed for that entity could be different. Moreover, some attack types could be observed more frequently compared to some other types within an entity. Therefore, the training and test data sets of each attack type could be more balanced or imbalanced depending on the chosen $t_g$ value. If a malicious email attack is observed a couple of times per week, then training a classifier with the ground truth data of each day (or 2 or 3 days) could be a better approach compared to training with the ground truth of a week. Similarly, if an attack type is seen each and every day consistently, training a model with attacks observed in 6 or 12 hours of granularity might be better. Based on this intuition, this work develops a general cyber attack forecasting framework for entities \ie organizations and industries that uses different $t_x$ and $t_g$ time granularities to forecast cyber attacks for different attack types. 

The infrequency of certain types of attacks may cause challenges where the positive instances are much fewer compared to the negative instances (no attack). Dealing with imbalanced data sets is challenging, because it is not often possible for a classifier to learn the minority class. There are several approaches to deal with imbalanced data sets. Apart from choosing a proper performance metric or classifier, other approaches include changing the distribution of the instances, which is a very delicate process. Moreover, there are many accompanying questions that need to be answered, like which instances to under or over sample and how to generate synthetic new instances? This work proposes a new approach to improve cyber attack forecasting performance on imbalanced data sets. The proposed algorithm is a modified version of the Synthetic Minority Over-sampling Technique (SMOTE) \cite{Chawla:2002}, where some of the majority instances are under sampled before applying SMOTE. We show that this algorithm improves the prediction performance of the proposed method on imbalanced data sets. 

The unconventional signals used to predict cyber attacks are not directly related to the target entity for which the prediction is made. There is no explicit relationship between these signals and the target entity. In fact, it is not easy to measure the level of correlation between any of the unconventional signals and the ground truth events. Bayesian networks are probabilistic graphical models that are more successful in marginalizing the unknown uncertainties compared to other models. They are helpful in accounting for the uncertainty among the variables whose distribution or dependencies are unknown and also extracting the influential or causal relationships among features if any. Therefore, Bayesian approach is used to predict cyber attacks with unconventional signals, as the relationship of these signals and the target entities is not straightforward.  

This paper is organized as follows: Section 2 presents a brief review of the previous approaches in cyber attack prediction. In Section 3, the problem is formulated by giving a background on Bayesian networks and defining the different time granularities used for signal and ground truth calculation. Section 4 explains the proposed approach, Section \ref{expres} gives the experiment results and Section \ref{concl} presents conclusions and future work.

\section{Previous Work}
Traditional intrusion detection systems rely on a misuse based approach where monitored events are matched against the signatures of the previously observed incidents \cite{Cannady98, Kruegel2003, Zhang2008, Li2012, Livadas2006}. One of the drawbacks of such systems is their inability to detect new events whose signatures are not known to the detection systems. Furthermore, deploying the signatures of the new attacks across the whole network takes some time and may not be effective. Anomaly detection aims at detecting deviations from normal behaviour and labels them as malicious \cite{Denning, Lippmann00, Blowers2014, Bilge2014, KruegelAndMutz2003}. These approaches typically have a high false alarm rate, because they may label normal but previously unseen behaviors as anomalies. 

Yang \textit{et al.} \cite{Yang2014} describe several attack projection frameworks which model how an attack might transpire over time. This modeling is broader than the traditional definition of the intrusion detection systems where observables of ongoing attacks are used to predict the next malicious actions based on system vulnerabilities and attacker behavior. In attack projection, the focus is on the footprints of the multistage cyber attacks. Extracted footprints and the sequence or causal dependencies among these footprints are used to generate hypothesized attack strategies that represent a multistage attack. These attack strategies are then modeled to project future actions of ongoing attacks. Wang \textit{et al.} \cite{Wang2006} suggest that efficient algorithms are needed for multi-step intrusions to correlate isolated alerts into attack scenarios and propose to use attack graphs. They use attack graphs to identify possible cases where vulnerabilities can be exploited in a network and show that their method is more successful in correlating isolated attacks into attack scenarios compared to the traditional intrusion detection systems. Similarly, Qin and Lee \cite{Qin} suggest to use probabilistic inference with a dynamic Bayesian Network approach to correlate attacks. Using the Grand Challenge Problem (GCP) data set of the Defense Advanced Research Projects Agency, they show that their method is successful in correlating isolated attacks, identifying attack strategies, and predicting future attacks. 

A broad range of Machine Learning techniques have been used so far for  intrusion detection and cyber attack prediction. Artificial Neural Networks \cite{Cannady98} \cite{Lippmann00}, Clustering \cite{Blowers2014}, Decision Trees \cite{Kruegel2003}\cite{Bilge2014}, Ensemble Learning \cite{Zhang2008}, Support Vector Machines \cite{Li2012} are some of the methods that have been used in the cyber security literature. Besides the traditional approaches, there are some studies that elaborate on the prediction of multi stage cyber attacks. For example, Cheng \textit{et al.} \cite{Cheng2011} use a novel method that is based on the Longest Common Subsequence (LCS) technique to measure the similarities between attack progressions, correlate security alerts with specific patterns, and predict multi stage attacks. They show that their method is able to project possible attacks better when compared to the previous LCS based approaches. Furthermore, Fava \textit{et al.} \cite{Fava2008} present the use of a variable-length Markov model (VLMM) that captures the sequential properties of attacks to project future activities of ongoing cyber attacks and show that the method they propose is able to adapt to the newly observed attack sequences.

In most of the existing intrusion detection or projection systems, the observables are direct outcomes of the malicious activities on the computing systems. These observables may be referred to as conventional signals. However, our research uses unconventional signals that are observed before any malicious activity occurs. They may be due to increasing negative tones toward an organization or news reports, and sometimes may not have explicit relations to the future victim of the cyber attacks. Preliminary studies have shown the viable concept of using unconventional signals to forecast cyber attacks \cite{Okutan2017} and attack intensities \cite{Werner2017}. Silver \cite{Silver} investigates how a true signal can be distinguished from a universe of noise data and states that most predictions fail due to a poor understanding of probability and uncertainty. He believes that as the appreciation of uncertainty improves, better prediction results could be achieved. Furthermore, Tetlock and Gardner \cite{Tetlock} state that creating good forecast involves gathering evidence from a variety of sources rather than using very powerful computing resources or arcane techniques.  

Bayesian methods have also been used in the cyber security literature to a limited extent \cite{KruegelAndMutz2003, Livadas2006}. Dua and Du \cite{Dua:2011} state that the false alarm rate of the rule based misuse detection systems are high since the signature of a malicious and normal user might overlap to activate a rule. As an alternative solution they suggest to use Bayesian networks and state that Bayesian approaches have a relative resilience in conditional probability table parametrization and allow the anomalousness of an event to be directly related to its probability. 

\section{Problem Formulation}
A Bayesian network is defined as a directed acyclic graph (DAG) that is composed of $n$ random variables (nodes) and $e$ edges that represent the conditional dependencies among these variables. Let $X = \left \{ X_1, X_2,...X_n \right \}$ be $n$ random variables (unconventional signals) with nominal or numeric values for a Bayesian network $B$ where $n >= 1$. Assuming $\pi_i$ represents the parents of node $X_i$, the probability distribution of $X_i$ is calculated by $P(X_i|\pi_i)$. Each node $X_i$ in the Bayesian network is associated with a conditional probability table that shows the probabilities for each value of $X_i$ based on all combinations of values of its parent nodes and is represented as $P_{X_i} = P(X_i|\pi_i)$. 

To calculate the joint probability distribution of $X$,  the chain rule is used \ie
\footnotesize
\begin{eqnarray}
P(X) &=& P(X_1|X_2,X_3,...,X_n)P(X_2,X_3,...,X_n) \nonumber \\
     &=& P(X_1|X_2,...,X_n)P(X_2|X_3,...,X_n)P(X_3,...,X_n) \nonumber \\
     &=& P(X_1|X_2,...,X_n)P(X_2|X_3,...,X_n)...P(X_{n-1}|X_n)P(X_n) \nonumber \\
     &=& \prod_{i=1}^{n} P(X_i|X_{i+1},...,X_n)
\end{eqnarray}
\normalsize
Knowing that $X_i$ is independent from the variables other than its parents $\pi_i$ 
\begin{equation}
P(X) = \prod_{i=1}^{n} P(X_i|\pi_i)
\end{equation} 
\begin{figure}[h!]
\centering
\includegraphics[scale=0.7]{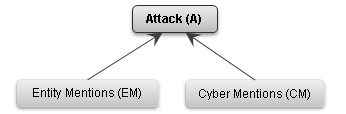}
\caption{An example Bayesian Network showing a hypothetical relationship among the Entity Mentions (EM), Cyber Mentions (CM) and Attack (A)}  
\label{fig:bn1}
\end{figure}
Consider the Bayesian network in Figure \ref{fig:bn1}. Bayesian inference can be used to calculate the probability of observing an attack towards an entity based on all possible values of Entity Mentions (EM) and Cyber Mentions (CM). For example, assuming that EM and CM can take values either Low (L) or High (H), 
\footnotesize
\begin{eqnarray*}
P(A|EM=H) &=& P(A|EM=H,CM=H)P(CM=H|EM=H)  \nonumber \\
           &+& P(A|EM=H,CM=L)P(CM=L|EM=H)           
\end{eqnarray*}
\normalsize
$P(CM=H|EM=H)$ is $P(CM=H)$  and $P(CM=L|EM=H)$ is $P(CM=L)$, because the variables $EM$ and $CM$ are independent. Then, 
\footnotesize
\begin{eqnarray*}
P(A|EM=H) &=& P(A|EM=H,CM=H)P(CM=H)  \nonumber \\
           &+& P(A|EM=H,CM=L)P(CM=L)
\end{eqnarray*}
\normalsize
Consider a set of target entities composed of organizations or industries $E = \left \{E_1, E_2, ... E_t\right \}$,  a set of cyber attack signals $X = \left \{X_1, X_2, ... X_n\right \}$, and a set of attack types $A = \left \{A_1, A_2, ... A_m\right \}$. For any given entity $E_i$, if the set of attack types is defined as $A(E_i)$ where $A(E_i) \subseteq A$ and the set of cyber attack signals is defined as $X(E_i)$ where $X(E_i) \subseteq X$, Bayesian inference is used to calculate the probability of each attack type $A_i \in A(E_i)$ based on the cyber attack signals in $X(E_i)$.
  
\subsection{Signal and Ground Truth Time Granularity}

Depending on how frequent cyber incidents are observed for an entity $E_i$, making predictions using different ground truth granularities could be helpful. If a couple of cyber attacks of a certain type are observed almost every day consistently, it would not be beneficial to train and test a classifier to predict attacks for each day. Instead, one may make predictions for 6 or 12 hours of granularity. On the other hand, depending on the availability, quality or update frequency of the unconventional signals, one may think to calculate signals for different time granularities \ie the last couple of days, weeks or months. If the data source is not being updated on a daily basis, it might not be beneficial to calculate signals on a daily basis. Different time granularities are used to calculate the unconventional signals and the ground truth. The time granularity of signals is defined as $t_x$ and five cases are considered where each signal is calculated by averaging its previous values observed in the last three days (3d), one week (1w), one month (1m), three months (3m), and six months (6m).

Similarly, the time granularity of the ground truth is defined as $t_g$ and four cases are evaluated where $t_g$ takes a value of 6, 12, 24, or 48 hours (The $t_g$ values are represented as 6hr, 12hr, 24hr, and 48hr respectively in the following sections). Taking $t_g$ less than 6 hours leads to a very sparse data set, since a typical organization may not see certain cyber attacks every hour of the day. Similarly, taking $t_g$ larger than 48 hours causes the training data set to have positive ground truth for the vast majority of instances. Depending on how frequent cyber incidents are observed for an entity or attack type, different $t_g$ values may be assessed for various entities and attack types.

For each attack type of an entity $E_i$, each of the $t_x$ values defined above is used to calculate the unconventional signals in $X(E_i)$. Similarly, to train a Bayesian classifier for each attack type, different $t_g$ time granularities (6, 12, 24, and 48 hours) are used to consider real cyber incidents for $E_i$. For example, if $t_x$ is selected as one week and $t_g$ as 24 hours, then for a given time $t$, signals calculated for the last week are used together with the ground truth of the next 24 hours ($t_g$) to train a Bayesian classifier. A Bayesian classifier is built for each attack type $A_i \in A(E_i)$ of a target entity $E_i$, considering all possible combinations of the defined $t_x$ and $t_g$ values.

\section{Proposed Method}
Bayesian inference is used to calculate the probability of cyber attacks towards a specific target entity using unconventional signals that may or may not be related to that target entity. These signals are pulled from various data sources, including social media (Twitter) and open source global event tracking projects like GDELT. 

The prediction is regarded as a classification problem where the value of the target class is either 1 or 0 depending on having a cyber attack or not. For each entity $E_i$ and its attack type $A_i \in A(E_i)$, the unconventional signals in $X(E_i)$ are used as random variables and the ground truth of $E_i$ for $A_i$ as target class to train a Bayesian classifier for different $t_x$ and $t_g$ pairs. For a given time $t$, the training set of each $A_i$ is represented as $S_{A_i}$ and includes the unconventional signals calculated for the time period $(t - t_x, t)$ and the real cyber incidents reported for the period $(t,t + t_g)$. For each attack type $A_i$ of a target entity $E_i$ and for each $t_x$ and $t_g$ time granularity pair, we learn the structure of the Bayesian network for $A_i$ using the previously observed unconventional signals in $(t - t_x, t)$ and the ground truth incidents observed in $(t,t + t_g)$. Then, for a given time $t$ unconventional signals are calculated for the time period $(t - t_x, t)$ to predict the cyber incidents for the upcoming time period \ie $(t,t + t_g)$. 

\subsection{Unconventional Signals Used}
\label{indicators}
Increased discussion of past or potential cyber attacks may be due to a significant change in the intent or capability of attackers or software vulnerabilities. Similarly, an increase in the number of mentions of a target entity might be due to an increase in the surveillance towards it, and may lead to an increase in the probability of a cyber attack for that target entity. Furthermore, the number of mentions of negative events, the tone of the negativity of these events or the number of source documents referring to them might be indicative of a cyber incident. These example unconventional signals are used to predict cyber attacks for an anonymized private company nicknamed KNOX. These unconventional signals are by no means comprehensive, but are good examples to test our methodology. These signals are described below, which are fed into a Bayesian classifier for each $t_x$ and $t_g$ time granularity pair, to predict attacks for each attack type defined for KNOX: Defacement, Malware, DOS, and Malicious Email/URL (MEU). 

\begin{itemize}
\item \textbf{Twitter Cyber Mentions (TCM)}: Given a time $t$ and a signal time granularity $t_x$, a set of cyber keywords including Hack, Malware, DDOS, and Malicious are used to count the number of mentions of these keywords on Twitter for each signal calculation period \ie $(t - t_x, t)$. 

\item \textbf{Twitter Entity Mentions (TEM)}: The number of mentions of KNOX related keywords in Twitter is counted for each signal calculation period $(t - t_x, t)$ to check if these mentions are indicative of a cyber incident towards KNOX.

\item \textbf{GDELT Event Mentions (GEM)}: GDELT keeps track of the media from every country, in over 100 languages. As a  measure of event significance it counts the total number of mentions of specific events across all of its source documents. GDELT also attaches an average tone value for each mentioned event. The value of the average tone could be between -100 and +100 and it indicates the degree of the negativity of the associated event. For each signal calculation period $(t - t_x, t)$, the total mentions of events that have a negative average tone are calculated. 

\item \textbf{GDELT Event Articles (GEA)}: GDELT counts the total number of source documents containing one or more mentions of an event for assessing the ``importance'' of an event. The more an event is discussed, the more likely it is regarded as significant. For a given time $t$, the total number of source documents mentioning events with an average negative tone are counted in the signal calculation period $(t - t_x,t)$. 

\item \textbf{GDELT Event Tone (GET)}: Given a time $t$ and a signal granularity $t_x$, the average tone of the negative events is calculated for the time period $(t - t_x,t)$.  

\end{itemize}

\subsection{Signal and Ground Truth Calculation}
For a given time $t$, the signals for the past period $(t - t_x, t)$ are aggregated to predict the cyber attacks for the next $t_g$ hours. Most of the signals are general and not specific to KNOX. The ground truth data of KNOX was available between April 1 and October 30 2016, therefore the prediction models are trained and tested in this period. Step by step details of the signal and ground truth generation process are provided in Algorithm \ref{signal_generation}. The variables $gtStartTime$ and $gtEndTime$ are set to 04.01.2016 and 10.30.2016 respectively. For each attack type and time granularity pair ($t_x$ and $t_g$), a separate data set is generated. All ground truth counts that are greater than 1 are set to 1 to perform binary prediction. In the inner most while loop, the unconventional signals are aggregated over a time period of length $t_x$ \ie starting at $signalStartTime$ and ending at $currentTime$. 

\begin{algorithm}[h!]
\caption{The signal and ground truth calculation process for different $t_x$ and $t_g$.}
\label{signal_generation}
\SetAlgoNoLine
\Fn{GenerateDataSets ($gtStartTime$, $gtEndTime$)}{
$attackTypes := [$Malware$, $Defacement$, $DOS$, $MEU$]$\;
$signalGranularity := [$6m$, $3m$, $1m$, $1w$, $3d$]$\;
$groundTruthGranularity := [$48hr$, $24hr$, $12hr$, $6hr$]$\;
\BlankLine
\ForEach{$a_i \in attackTypes$}{%
\ForEach{$t_x \in signalGranularity$}{%
\ForEach{$t_g \in groundTruthGranularity$}{%
	$currentTime := gtStartTime$\;
	\While{$currentTime <= gtEndTime - t_g$}{
		$signalStartTime := currentTime - t_x$\;
  				
		\BlankLine
		\Comment{Calculate average of signals TCM, TEM, GEM, GEA, GET}
		let $signals$ be the signals calculated for time period $(signalStartTime, currentTime)$\;
		
		\BlankLine
		\Comment{Calculate ground truth}
		$gtEnd := currentTime + t_g$\;
		let $gt$ be number of cyber incidents observed in the time period $(currentTime, gtEnd)$\;
		\If{$gt > 1$}{
			$gt := 1$\;
		}
		
		\Comment{signal and gt is appended to the file for $a_i$}
		$write(a_i, signals, gt)$\;
		
		\BlankLine
		$currentTime := currentTime + t_g$\;
	}
}
}
}
}
\end{algorithm}

\subsection{Working with Imbalanced Cyber Data Sets}
Cyber incidents towards a specific entity are rare events that can be marked as anomalies especially when filtered for a specific attack type. Therefore, the data sets for cyber incidents could be highly skewed towards negative instances that have a ground truth value of zero. Sometimes it might be very difficult for a classifier to predict positive instances with a cyber attack based on such data sets. There are several approaches to overcome the imbalanced data set problem, including:
\begin{itemize}
\item Using different performance metrics like F-Measure or AUC (the area under the ROC curve) rather than accuracy.
\item Using different classifiers like decision tree based algorithms that could perform well on imbalanced data sets.
\item Over sampling the minority class. 
\item Under sampling the majority class.
\item Introducing new synthetic instances for the minority class.
\end{itemize} 
A new algorithm named SMOTE++ is proposed to improve the cyber attack prediction performance on imbalanced data sets. It is a hybrid approach where the under sampling, instance weighing, and over sampling techniques are used together.

It is highly desirable for a classifier to have the negative and positive instances apart from each other. In other words, if it is easier to separate the negative and positive instances with a line, plane or hyperplane, one may expect a classifier to perform better. Unconventional signals that are not explicitly related to the target entity are used. Therefore, some of the training instances may not be sufficient in explaining the relationship of the unconventional signals and the cyber event ground truth. To be able to distinguish the negative and positive instances better, we suggest to remove a certain percentage of the majority class (negative) instances that are near to the minority class (positive) instances. To find such instances, a simple approach could be to filter out the negative instances that are near to the mean of the positive instances. $k$-Means clustering algorithm is used to find the main cluster of the positive instances and remove some portion of the negative instances that are near to that mean where the percentage to remove is a configurable parameter in the algorithm. In case there is no cluster for the positive instances \ie they are scattered around, the mean of all positive (minority) instances is used to remove negative (majority) instances. Removing some of the negative instances causes a change in the total weight of the negative classes. To compensate for that, the remaining negative instances are reweighed to maintain the same total weight for the negative instances. On the other hand, to make the class weights of the negative and positive classes equal, a hybrid approach is followed where the weights of the existing positive instances are reweighed and new synthetic positive instances with a lower weight are introduced. This is done to differentiate the existing minority instances from the newly genareted synthetic ones. Assume that we need to introduce three more instances for each positive instance to achieve an equal weight for the negative and positive classes. We first multiply the weight of each existing positive instance by two and then introduce two new synthetic instances for every positive instance. To introduce new synthetic instances $k-NN$ algorithm is used in the same way it is used in SMOTE \cite{Chawla:2002}. A step by step description of the proposed method is shown in Algorithm \ref{algSMT}.

\begin{algorithm}[h!]
\caption{SMOTE++ algorithm.}
\label{algSMT}
\SetAlgoNoLine
\SetKwInOut{Input}{Input}
\SetKwInOut{Output}{Output}
\Input{An imbalanced data set denoted as $allInstances$ \newline 
	   Percentage of majority instances to remove \ie $p$ \newline
	   The number of nearest neighbors to consider to generate synthetic minority instances \ie $k2$}
\Output{A new data set with a uniform majority and minority class distribution}

\BlankLine
\Fn{SMOTE++ ($allInstances$, $p$, $k2$)}{
	let $majInstances$ be the set of majority instances in $allInstances$\;
	let $minInstances$ be the set of minority instances in $allInstances$\;
	\BlankLine
	let $sMin$ be the size of the $minInstances$\;	
	let $sMaj$ be the size of the $majInstances$\;
	
	\BlankLine
	\Comment{Find the first minority cluster using $k-$Means Clustering with Euclidian distance}
	$k := 2$\;
	$minorityClusterFound := false$\;	
	\While{$minorityClusterFound \neq true$}{ 
		let $clusters$ be the first $k$ clusters in $allInstances$\;
		\uIf{$clusters$ includes a minority cluster}{
    		let $cMin$ be the centroid of the minority cluster in $clusters$\;
    		$minorityClusterFound := true$\;
  		}
  		\Else{
    		$k := k + 1$\;
  		}	
  		\uIf{$k = sMin$}{
  			break\;
  		}
	}
	
	\Comment{If a minority class cluster is not found then use the mean of all minority instances}
	\BlankLine
	\uIf{$minorityClusterFound \neq true$}{
    		let $cMin$ be the mean of all minority instances in $minInstances$\; 		
  	}
  		
	\BlankLine
	\Comment{Filter majority instances}
	remove $p$ percent of $majInstances$ that are nearest to $cMin$\;
	let $majInstancesNew$ be the remaining instances in $majInstances$\;
	
	\BlankLine
	\Comment{Reweigh majority instances}
	$majWeight := 100 / (100 - p)$\;
	set the weight of each instance in $majInstancesNew$ to $majWeight$\;
	
	\BlankLine
	\Comment{Reweigh existing minority instances}		
	$minW := sMaj / sMin$ / 2\;
	set weight of each instance in $minInstances$ to $minW$\;

	\BlankLine
	\Comment{Generate new synthetic minority instances}
	generate $minW * sMin$ synthetic minority instances using $k-$NN with $ k := k2$ \cite{Chawla:2002}\;
	let $minInstancesSyn$ be the set of created synthetic minority instances\;
	\BlankLine
	\Return $majInstancesNew \cup minInstances \cup minInstancesSyn$\;
}

\end{algorithm}

\section{Experiments And Results}
\label{expres}
The unconventional signals defined in Section \ref{indicators} that are extracted from various data sources and are not explicitly related with KNOX are used to predict the probability of different attack types for KNOX. For each attack type, a separate data set is created for every $t_x$ and $t_g$ time granularity pair. Using five $t_x$ and four $t_g$ time granularity values, a total of 20 different data sets are created for each attack type. It is always desirable for a classifier to have a high true positive (TP) and a low false positive (FP) rate at the same time. The area under the ROC curve (AUC) gets higher, when TP is high and FP is low. Therefore, AUC is used to compare the performance of the proposed model on different data sets. Moreover, the BayesNet classifier in Weka \cite{Weka2009} is used with $10\times10$ folds cross validation to calculate the average AUC value for each attack type and the $t_x$ and $t_g$ time granularity pair.  

\subsection{Data Sets Used}
We create a separate data set for each attack type and the $t_x$ and $t_g$ time granularity pair. The distribution of the negative and positive classes in these data sets are different for different $t_g$ values. For each attack type, as $t_g$ gets lower, fewer ground truth incidents are observed for KNOX, therefore the data sets become more imbalanced. Similarly, as $t_g$ gets larger more cyber incidents are observed and the ratio of the positive instances with a cyber attack increases. The list of the created data sets and their negative and positive class distribution for each attack type and $t_g$ are shown in Table \ref{data_sets}. For example, the data sets for the DOS attack type are highly imbalanced for all $t_g$ values. However, the data sets for the Malware attack type are more balanced compared to the DOS and MEU attack types. Moreover, the data sets of the MEU and Defacement attack types are skewed when $t_g$ is low.
\begin{table}[!h]
\centering
\caption{The percentage (\%) of the positive instances for the data sets of different attack types for different $t_g$ time granularities: 6, 12, 24, and 48 hours. A positive instance corresponds to an instance with a cyber attack.}
\label{data_sets}
\begin{tabular}{lcccc}
 \hline
                    & \textbf{6 hr} & \textbf{12 hr} & \textbf{24 hr} & \textbf{48 hr} \\
                     \hline
\textbf{Malware}    & 36          & 51           & 72           & 80           \\
\textbf{Defacement} & 15          & 26           & 48           & 64           \\
\textbf{DOS} 		& 2 	      & 4  	         & 9  	        & 15           \\
\textbf{MEU}        & 10          & 17           & 32           & 50           
\end{tabular}
\end{table}

\subsection{Using Different $t_x$ and $t_g$ Time Granularities}
\begin{figure}[h!]
\centering
\includegraphics[scale=0.36]{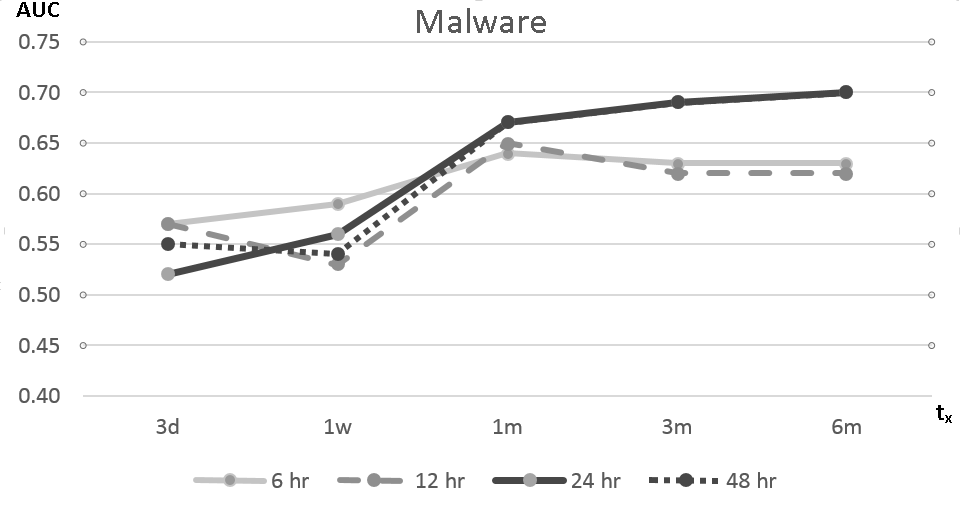}
\caption{Average AUC values for the Malware attack type for different $t_x$ and $t_g$ signal and ground truth time granularities.}  
\label{fig:Malware}
\end{figure}

\begin{figure}[h!]
\centering
\includegraphics[scale=0.36]{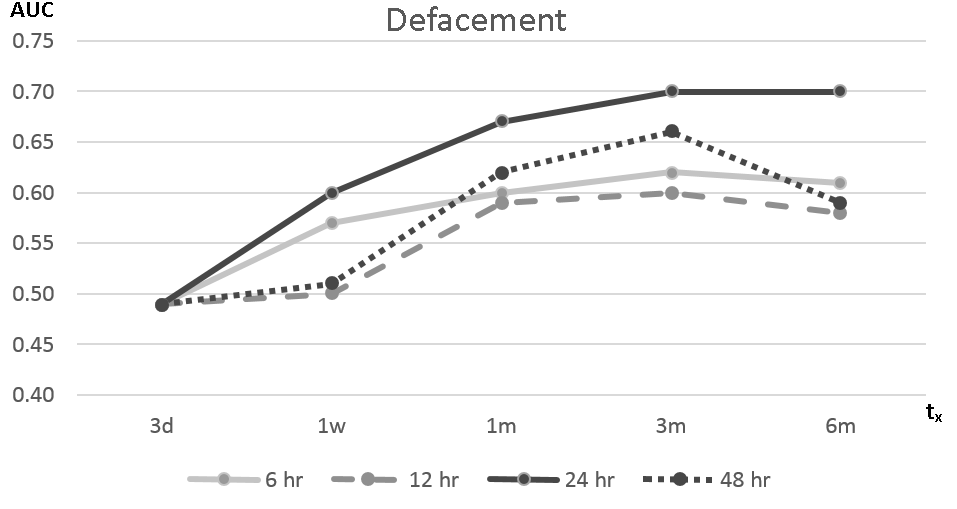}
\caption{Average AUC values for the Defacement attack type for different $t_x$ and $t_g$ signal and ground truth time granularities.}  
\label{fig:Def}
\end{figure}

\begin{figure}[h!]
\centering
\includegraphics[scale=0.36]{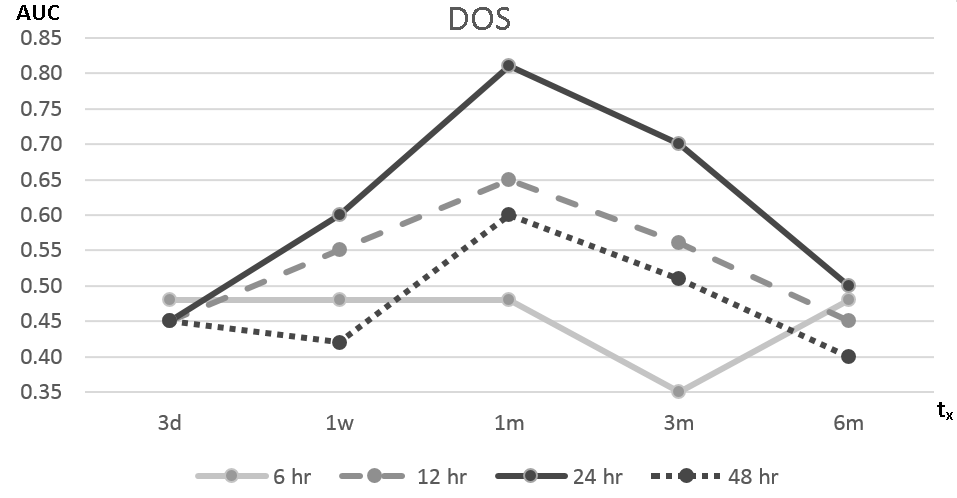}
\caption{Average AUC values for the DOS attack type for different $t_x$ and $t_g$ signal and ground truth time granularities.}  
\label{fig:dos}
\end{figure}

\begin{figure}[h!]
\centering
\includegraphics[scale=0.36]{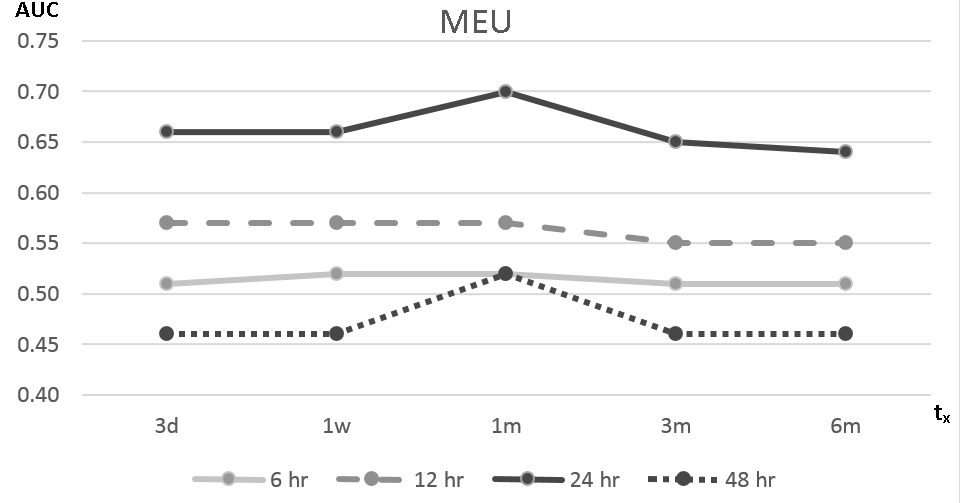}
\caption{Average AUC values for the MEU attack type for different $t_x$ and $t_g$ signal and ground truth time granularities.}  
\label{fig:MEU}
\end{figure}

The average AUC values for each attack type for different $t_x$ and $t_g$ pairs are shown in Figures \ref{fig:Malware}, \ref{fig:Def}, \ref{fig:dos}, and \ref{fig:MEU}. We present the AUC values when $t_x$ is three days (3d), one week (1w), one month (1m), three months (3m), and six months (6m) for different $t_g$. Each line in these figures shows the AUC values when $t_g$ is 6, 12, 24, and 48 hours respectively. First, we observe that when an optimum $t_x$ and $t_g$ time granularity is chosen, it is possible to achieve an AUC value of at least 0.70 for all attack types defined for KNOX, using unconventional signals that are not directly related with KNOX. The results suggest that if a prediction model calculates signals for different time granularities and keeps track of the best performing granularity pair, it might be possible to achieve a better performance to predict cyber incidents for different attack types. 

A data set might be more imbalanced for specific attack types (like DOS or MEU in our case) or when created with a lower $t_g$ value. For example, when $t_g$ is 24 hours, the rate of the positive instances with a cyber attack and the negative instances with no attack are different for each attack type. One might expect a lower performance on an imbalanced data set compared to a data set that has a uniform class distribution. Although the skewness of data sets are different for each attack type when $t_g$ is 24 hours, a higher classification performance is observed for all attack types for $t_g=24$ hours. For example, except for Malware, the performance of the prediction model is better when $t_g$ is 24 hours for almost all attack types and $t_x$ values. For Malware, a comparable performance is observed when $t_g$ is 24 and 48 hours. 


Similarly, the performance of the classifier is also impacted by the choice of $t_x$. For example, for the MEU and DOS attack types the AUC value of the classifier is higher when $t_x$ is one month. However, Malware and Defacement follow a different pattern. For the Defacement attack type, a higher AUC value is observed when $t_x$ is three months and the performance is comparable when $t_x=6$ months. Similarly, for the Malware data set, although the performance of the model is slightly better when $t_x$ is six months, a comparable performance is observed when $t_x$ is three months.  

\subsection{Relationships of the Signals and Cyber Incidents}
To explore the relationship of each unconventional signal and the cyber incidents, we take a closer look at the Bayesian networks generated. Twenty Bayesian networks are trained for each attack type considering each possible pair of defined $t_x$ and $t_g$ values. One of the best performing granularity pairs is used for each attack type where $t_g$ is 24 hours and $t_x$ is either one month or three months depending on the average AUC value found. Therefore, the Bayesian network for $t_x=1m$ is used for the DOS and MEU attack types, and the Bayesian network for $t_x=3m$ is used for the Malware and Defacement attack types.  

Using a Bayesian network, assuming a dependency exists between a signal and the target class, we can find out the probability of having a high or low signal value when an attack is observed or not. Furthermore, by looking at the probability values in the conditional probability tables for the low and high values of each signal, one may comment on the importance of a specific signal. If a signal has a relationship with the target class and the probabilities for its high and low values are different for cases with and without a cyber attack, that signal might be regarded as important. Similarly, if the probabilities for its high and low values are close to each other for cases with and without a cyber attack, it has less discriminative power. For a signal to be more discriminative, if the probability of observing its high value is higher when a cyber attack is seen, it is expected to be lower when there is no cyber attack.        

\begin{figure*}[h!]
\centering
\includegraphics[scale=0.41]{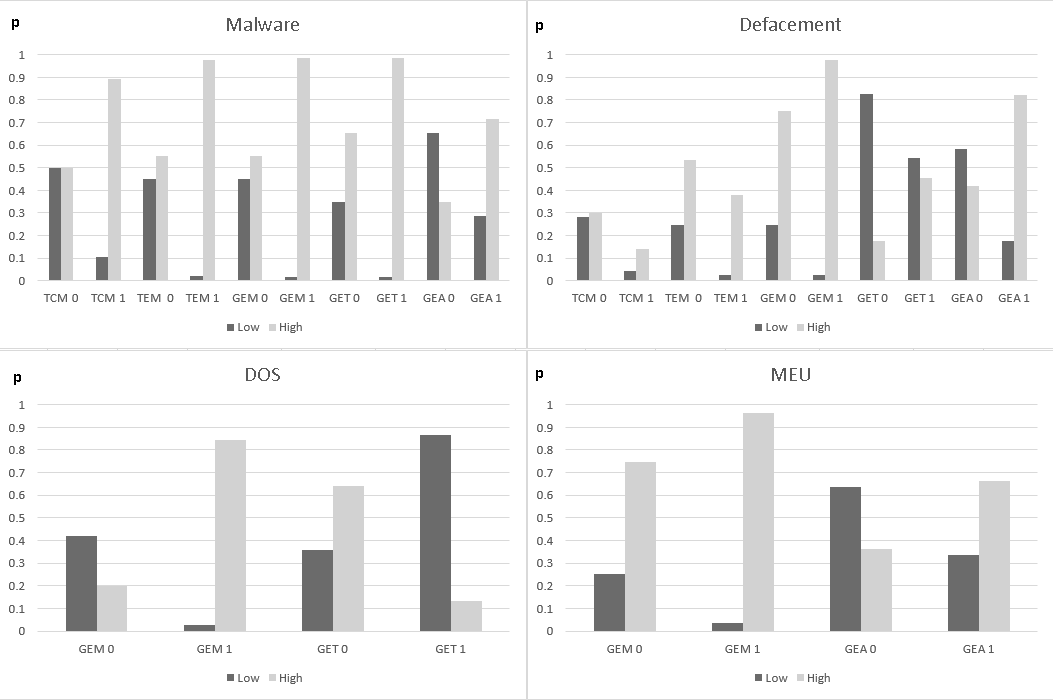}
\caption{The probability ($p$) of having a low or high value of a signal when there is an attack or not. All probabilities in each graph are based on the conditional probability tables of the Bayesian networks learned for the corresponding attack types}  
\label{fig:all_cpt}
\end{figure*}

We look at the conditional probability tables of each signal in the Bayesian networks learned for each attack type. For the Malware attack type, a relationship is observed between each unconventional signal and the cyber attacks. The values in the conditional probability table when $t_x$ is 3 months and $t_g$ is 24 hours is shown in Figure \ref{fig:all_cpt}. When an attack is observed, the probability of having a higher value of TCM, TEM, GEM, GET, and GEA signal is much higher. However, for the TCM signal, when there is no attack, the probability of having a high and low signal value is equal, therefore TCM is less predictive compared to other signals when the ground truth is zero. On the other hand, GEA tends to be lower when the ground truth is zero and higher when the ground truth is one, therefore it is more discriminative for the Malware attack type compared to other signals.   

The relationship of the signals and the target class for the Defacement, DOS, and MEU attack types is shown in the corresponding graphs in Figure \ref{fig:all_cpt}. The $t_g$ time granularity is 24 hours for each of the data sets used for these attack types. The $t_x$ time granularity is one month for the DOS, MEU, and 3 months for the Defacement attack type. Only the unconventional signals that have a relationship with the target class are included in the graphs. We find that when a Defacement attack is observed, the value of the TCM, TEM, GEM, and GEA signals tend to be higher. For the DOS and MEU attack types, there are fewer signals that have a relationship with the target attack class. For the DOS attack type, GEM is more discriminative compared to the GET signal, as its probabilities for the low and high signal values are very much different when the ground truth is 0 and 1. A similar case is observed for the MEU attack type with the GEA signal. GEA tends to have a lower value when the ground truth is zero, and a higher value when the ground truth is one. Although GEM is marked as an important signal for the MEU attack type, it is less discriminative compared to GEA. The analysis above provides a means to assess the quality of the signals to discriminate positive and negative instances. This is made possible with the use of Bayesian networks. Similar analyses with additional unconventional signals could be helpful to explore other important signals for different attack types.  
 
\begin{figure}[h!]
\centering
\includegraphics[scale=0.35]{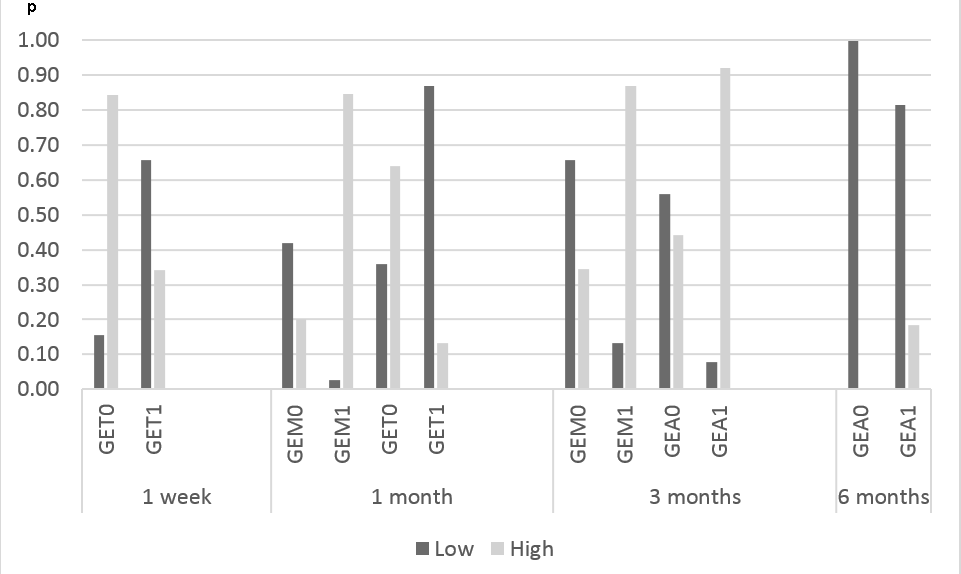}
\caption{The probability ($p$) of observing a DOS attack depending on the high or low values of the unconventional signals for different $t_x$ values when $t_g$ is 24 hours.}  
\label{fig:dos_all_t1}
\end{figure}

For a given attack type, having a higher AUC value for a specific $t_x$ may not mean that all unconventional signals are more discriminative for that $t_x$. The importance or discriminative power of a signal might also be changing for different $t_x$ values. We use the DOS attack type as an example, to show how the quality or importance of each signal changes with different $t_x$. The probabilities of the high or low values of each signal is shown in Figure \ref{fig:dos_all_t1}. GEA signal seems to be important when $t_x$ is three and six months and it is more discriminative when $t_x$ is three months. However, GEA is not among the signals that have a dependency with the target class when $t_x$ is one month. Therefore, we use the GEA signal calculated for $t_x = 3m$ and other signals calculated for $t_x = 1m$, to repeat the experiment for $t_x = 1m$. As a result, the AUC value of the BayesNet classifier increases from 0.806 to 0.824 for $t_x = 1m$, when GEA signal calculated with $t_x = 3m$ is used (See Table \ref{tbl_pick_diff_tx}).  

\begin{figure*}[h!]
\centering
\includegraphics[scale=0.45]{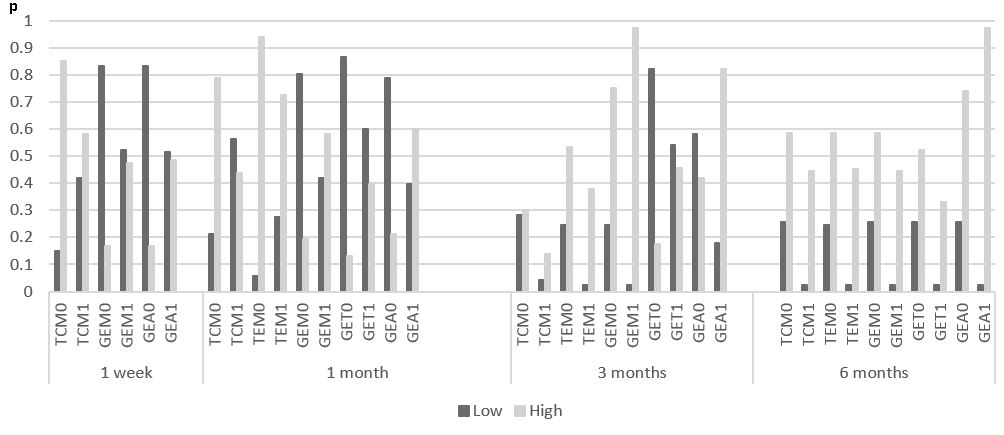}
\caption{The probability ($p$) of observing a Defacement attack depending on the high or low values of the unconventional signals for different $t_x$ values when $t_g$ is 24 hours.}  
\label{fig:def_all_t1}
\end{figure*}

Similarly, the quality of each signal calculated with different $t_x$ granularities is analyzed for other attack types. For example, when $t_x = 3m$ and $t_g = 24hr$ the AUC value is 0.696 for the Defacement attack type (See Figure \ref{fig:Def}). The conditional probability tables for each signal in the Bayesian networks learned for Defacement for different $t_x$ show that the GEM signal is more powerful when $t_x = 1m$ (See Figure \ref{fig:def_all_t1}). Therefore, the GEM signal (calculated with $t_x = 1m$) is used together with the remaining signals calculated with $t_x = 3m$ and this increases the AUC value from 0.696 to 0.714. On the other hand, for the Malware and MEU attack types, all signals that have a dependency with the target class seem to be more predictive when $t_x = 3m$ and $t_x = 1m$ respectively. The AUC values found when picking up a more predictive signal with a different $t_x$ is shown in the ``\textbf{Variable $t_x$}'' column of Table \ref{tbl_pick_diff_tx} for each attack type. We find that depending on an entity or attack type, there might be a benefit if more discriminative signals with different $t_x$ are used to built the Bayesian prediction model.             

\begin{table}[h!]
\centering
\caption{The AUC values for the Defacement and DOS when using signals with fixed and variable $t_x$}
\label{tbl_pick_diff_tx}
\begin{tabular}{lll}
\hline
   & \textbf{Fixed $t_x$} & \textbf{Variable $t_x$} \\
\hline
Defacement & 0.696        & 0.714        \\
DOS & 0.806        & 0.824       
\end{tabular}
\end{table}

To cross check our findings about the importance of each unconventional signal for each attack type, feature selection technique is applied to the same data sets with $10\times10$ folds cross validation. The CfsSubsetEval algorithm \cite{Hall1998} is used with the BestFirst search method to select more predictive signals in each data set. CfsSubsetEval selects the subsets of signals that are highly correlated with the class in each fold. The results for the CfsSubsetEval is shown in Table \ref{tbl_feature_sel}. For each attack type, the number of folds a signal is selected by CfsSubsetEval is shown for each signal. For the Malware attack type, the GEM signal is selected in all folds and the GET, TEM, and GEA signals are selected in at least one fold. Similarly, signals that are differentiative according to the conditional probability tables of the Bayesian classifier are selected by the CfsSubsetEval for other attack types. For example, for the DOS attack type only the GEM and GET signals are selected. Similarly, CfsSubseteval selects the GEM and GEA signals for the MEU attack type. As a summary, we show that the feature selection technique confirms our previous observations with the conditional probability tables, with regards to the importance of the unconventional signals.  

\begin{table}[]
\centering
\caption{The number of times each signal is selected in the 10-fold cross validation experiment of each attack type.}
\label{tbl_feature_sel}
\begin{tabular}{llllll}
\hline
           & \textbf{TCM}        & \textbf{TEM} & \textbf{GEM} & \textbf{GET} & \textbf{GEA} \\ \hline
\textbf{Malware}    & 0   & 1            & 10           & 2            & 1            \\
\textbf{Defacement} & 9   & 10           & 4            & 5            & 10           \\
\textbf{DOS}        & 0   & 1            & 10           & 10           & 0            \\
\textbf{MEU}        & 0   & 0            & 8            & 0            & 3            \end{tabular}
\end{table}

\subsection{Working with Imbalanced Cyber Data Sets}
As $t_g$ gets lower, the percentage of the positive instances decreases in all data sets. For instance, when $t_g$ is 6 hours, only 2\% and 10\% of the instances are positive in the data sets for the DOS and MEU attack types. In fact the DOS data set is still highly imbalanced when $t_g$ is 48 hours. Observation of a cyber incident is an abnormal situation for a target entity, therefore most of the time the cyber data sets are highly imbalanced. There might also be cases where one needs to use a smaller $t_g$. In addition to DOS, which is inherently imbalanced, other attack types may also suffer from the imbalanced ground truth for smaller $t_g$ values. It is always more difficult for a classifier to classify instances in a highly imbalanced data set. Notice the poor performance of the Bayesian classifier (lower AUC values) for lower $t_g$. For instance, when $t_g$ is 6 hours, a lower AUC value is observed for the skewed Defacement, DOS, and MEU data sets in the Figures \ref{fig:Def}, \ref{fig:dos}, and \ref{fig:MEU}. Therefore, any technique that has a potential to improve the performance of a classifier on the imbalanced data sets is valuable. We use the widely used techniques in the literature together with the proposed SMOTE++ method, to check if the performance of the Bayesian classifier gets better on the imbalanced data sets. Unless a data set has a completely uniform class distribution it can be regarded as imbalanced, therefore existing filtering techniques and SMOTE++ are applied to all data sets for all attack types. Some of the widely used approaches to combat the imbalanced data set problem are:

\begin{itemize}
\item \textbf{SMOTE:} This technique generates synthetic minority instances using $k-NN$ algorithm \cite{Chawla:2002}. One can specify the percentage of the minority instances to generate, to convert a highly imbalanced data set to a completely balanced one. During our experiments, we choose the proper percentage parameter for each data set to make the distribution of the positive and negative classes equal.
\item\textbf{Spread Sub Sample:} This technique randomly sub samples the majority class to decrease the number of the majority instances. During our experiments, the distribution parameter of the method is chosen to be 1, to achieve an equal positive and negative class distribution in the data sets.
\end{itemize}
\begin{figure}[h!]
\centering
\includegraphics[scale=0.45]{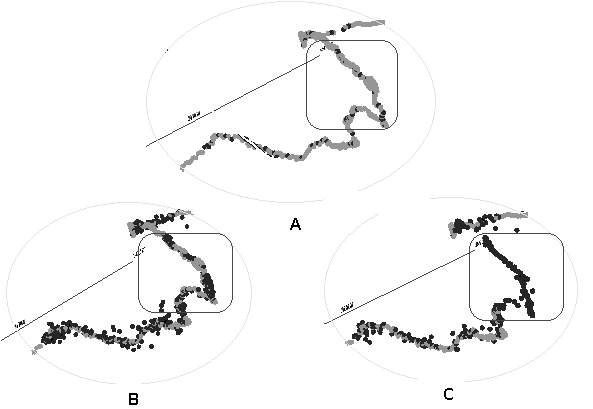}
\caption{A 2-D view of the data set for the MEU attack type before and after the SMOTE and SMOTE++ filters are applied. The plot of the MEU data set before any filters are applied is shown in (A). (B) shows the data set after the SMOTE filter is applied, and (C) shows the data set after the SMOTE++ filter is applied.}  
\label{fig:meu_smt}
\end{figure}

In addition to the existing filtering approaches, SMOTE++ is used to achieve a uniform class distribution before applying the Bayesian classifier. In the SMOTE++ approach, a certain percentage of the majority instances that are nearest to the minority instances are removed. To visualize the changes made on a data set by SMOTE++, the scattered view of the MEU data set instances is shown in Figure \ref{fig:meu_smt}.A, before any filter is applied (when $t_x$ is one month and $t_g$ is 6 hours). Then, the scattered view of the same data set is shown after the SMOTE filter is applied in Figure \ref{fig:meu_smt}.B. Similarly, the status of the data is shown in Figure \ref{fig:meu_smt}.C, after the SMOTE++ technique is applied.

\begin{figure*}[h!]
\centering
\includegraphics[scale=0.45]{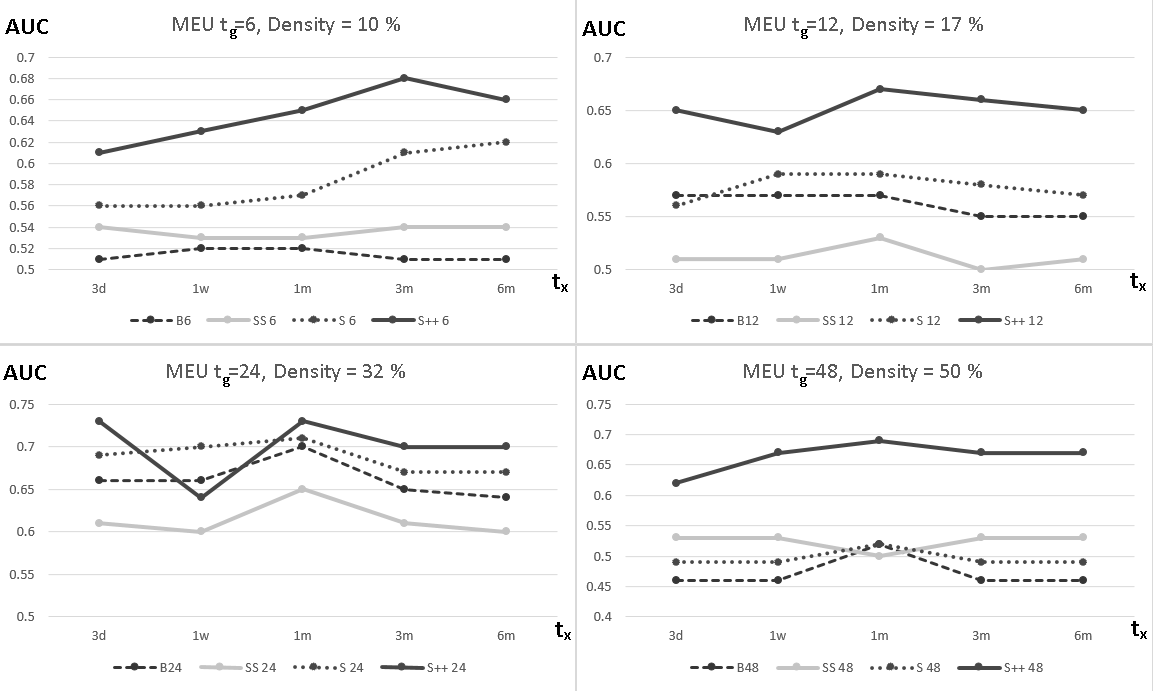}
\caption{Average AUC values for the data sets of the MEU attack type with different $t_x$ and $t_g$ values when SMOTE++ (S++), RandomSubsample (SS), and SMOTE (S) filters are applied before BayesNet classifier. The dashed line shows the AUC values with an ordinary BayesNet classifier (B) and the Density shows the percentage of positive instances in each data set.}  
\label{fig:meu_smtcmp}
\end{figure*}

We observe that the number of positive instances (dark ones) increased after the SMOTE filter is applied, compared to the original data set shown in Figure \ref{fig:meu_smt}.A. A similar situation is observed for the data set when SMOTE++ is applied too. But, an important difference exists between the two approaches when the highlighted parts of Figure \ref{fig:meu_smt}.B and \ref{fig:meu_smt}.C are considered. With SMOTE, the generated synthetic positive instances are mixed with the existing negative majority instances. However, when SMOTE++ is applied, the main cluster of the positive instances does not include that many negative majority instances. 
\begin{figure*}[h!]
\centering
\includegraphics[scale=0.45]{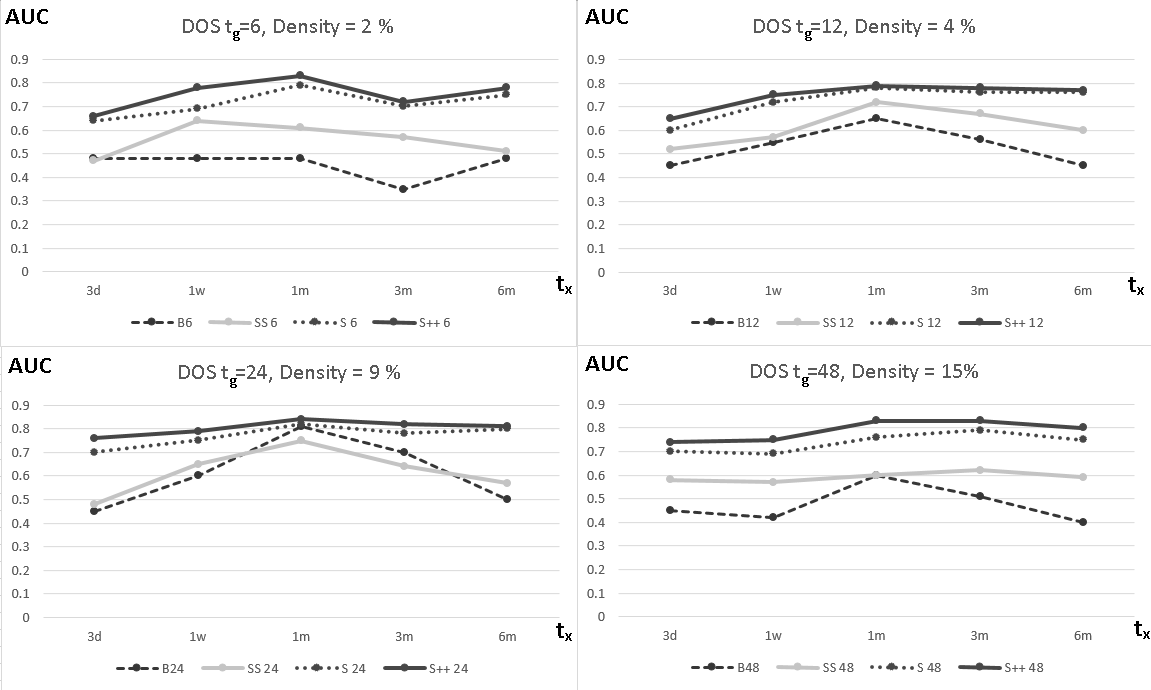}
\caption{Average AUC values for the data sets of the DOS attack type with different $t_x$ and $t_g$ values when SMOTE++ (S++), RandomSubsample (SS), and SMOTE (S) filters are applied before BayesNet classifier. The dashed line shows the AUC values with an ordinary BayesNet classifier (B) and the Density shows the percentage of positive instances in each data set.}  
\label{fig:dos_smtcmp}
\end{figure*}

\begin{figure*}[h!]
\centering
\includegraphics[scale=0.45]{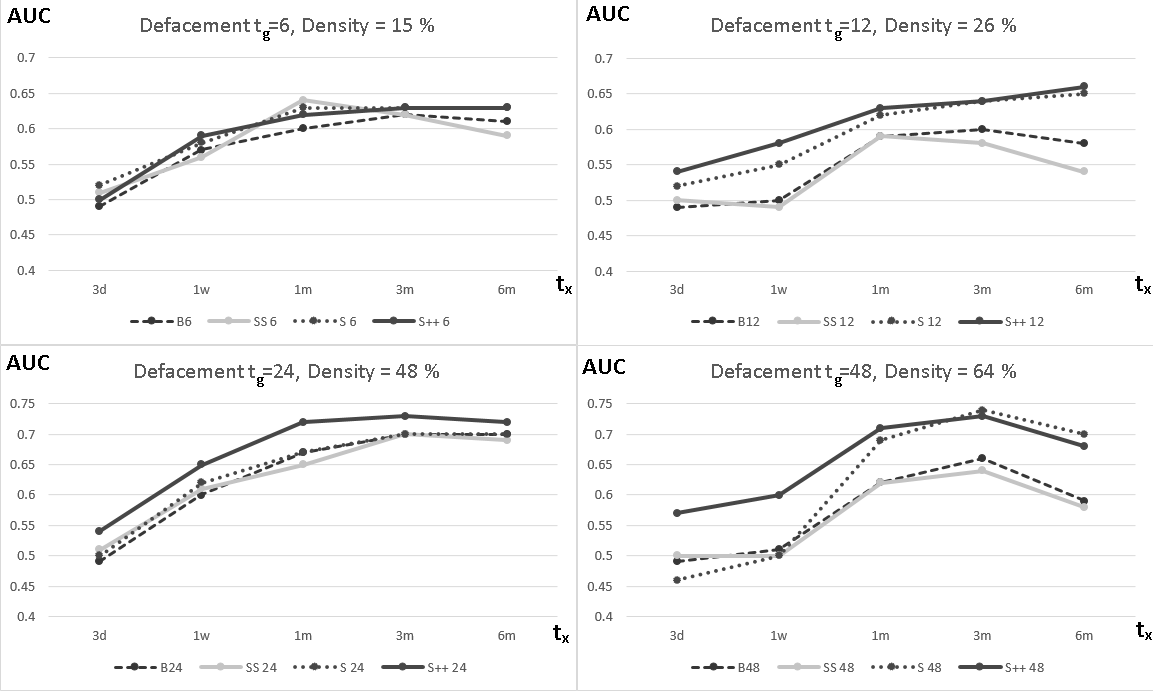}
\caption{Average AUC values for the data sets of the Defacement attack type with different $t_x$ and $t_g$ values when SMOTE++ (S++), RandomSubsample (SS), and SMOTE (S) filters are applied before BayesNet classifier. The dashed line shows the AUC values with an ordinary BayesNet classifier (B) and the Density shows the percentage of positive instances in each data set.}  
\label{fig:def_smtcmp}
\end{figure*}

\begin{figure*}[h!]
\centering
\includegraphics[scale=0.45]{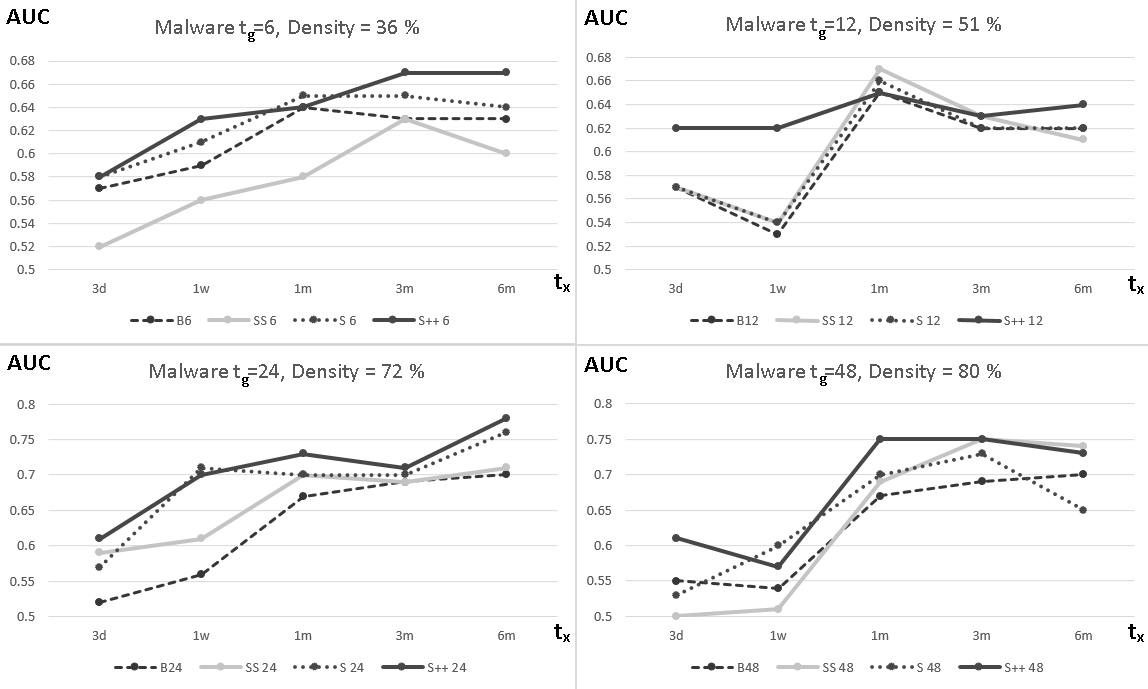}
\caption{Average AUC values for the data sets of the Malware attack type with different $t_x$ and $t_g$ values when SMOTE++ (S++), RandomSubsample (SS), and SMOTE (S) filters are applied before BayesNet classifier. The dashed line shows the AUC values with an ordinary BayesNet classifier (B) and the Density shows the percentage of positive instances in each data set.}  
\label{fig:malware_smtcmp}
\end{figure*}

Does cleaning the main cluster of the positive (minority) instances from the negative (majority) ones help the Bayesian classifier to separate the positive and negative instances better? SMOTE++ is applied in parallel with other filtering techniques in the literature, to see if the performance of the Bayesian classifier is improved. During each filtering, the BayesNet classifier is used with the FilteredClassifier algorithm in Weka \cite{Weka2009} to ensure that the test set is not touched when a filter is applied on each data set. Furthermore, the percentage parameter is tuned before applying SMOTE++ to a data set. The average AUC values of different methods, for different attack types and $t_x$ and $t_g$ values are shown in Figures \ref{fig:meu_smtcmp}, \ref{fig:dos_smtcmp}, \ref{fig:def_smtcmp}, and \ref{fig:malware_smtcmp}. The AUC values of the SMOTE++, SMOTE, and the RandomSubsample are shown with dark solid (blue), light solid (green) and dotted (blue) lines respectively. Additionally, the AUC values of the BayesNet without any filters is shown with a dashed (red) line in Figures \ref{fig:meu_smtcmp}, \ref{fig:dos_smtcmp}, \ref{fig:def_smtcmp}, and  \ref{fig:malware_smtcmp}. 

SMOTE++ performs better than other techniques in terms of AUC in almost all cases for all attack types and all $t_x$ and $t_g$ time granularities. To compare different methods and check if a difference of AUC between these methods is significant or not, a two tailed $t$-test is used with a $p$-value of 0.05 in all experiments. For the MEU attack type, the AUC value of the BayesNet is around 0.5 when $t_g$ is 6, 12, or 48 hours before any filter is applied. Although we do not observe a consistent and significant improvement with other filtering techniques, using SMOTE++ a significant increase is observed in the average AUC value of the BayesNet classifier when $t_g$ is 6, 12, or 48 hours.
For the DOS attack type, SMOTE++ achieves a better performance compared to SMOTE for all $t_x$ and $t_g$ values, but the differences in the AUC are not significant. Both methods lead to a significant improvement in the AUC value of the Bayesian classifier for all $t_g$ values and for almost all $t_x$ except when $t_x=1m$ and $t_g=24hr$. Although the RandomSubsample method improves the prediction performance of the Bayesian classifier in some cases, in some other cases its AUC value is comparable with the AUC value of the ordinary Bayesian classifier (See Figure \ref{fig:dos_smtcmp}).  

For the Defacement attack type, SMOTE++ achieves the best AUC scores in almost all cases compared to other filtering techniques but the differences are not significant except the cases when $t_g=48hr$ and $t_x$ is three days or one week. When compared to the Bayesian classifier with no filtering, SMOTE and SMOTE++ leads to a significant improvement in the AUC values when $t_g$ is 12 and 48 hours (See Figure \ref{fig:def_smtcmp}). Similar to our findings for other attack types, SMOTE++ achieves the highest AUC for the Malware attack type with most of the $t_x$ and $t_g$ combinations. For some data sets like when $t_x$ is one week and $t_g$ is 12 or 24 hours, the improvement is significant compared to the ordinary BayesNet classifier as it is seen in Figure \ref{fig:malware_smtcmp}. 
\section{Typical Risks for Research Validity}
In research studies there are three types of validity risks \ie internal, construct, and external that should be considered. Below we explain the measures taken to mitigate each of these validity risks.

If causal relationships between a dependent target and its independent variables are not defined properly in an experiment, then its internal validity is low. Although we use unconventional signals that are not directly related to the target entity, we still observe a similar and encouraging performance for all attack types with different skewness. Observing similar promising results for all attack types with different skewness and ground truth is important to mitigate the internal validity threats. In addition to that, using Bayesian networks clarifies the probabilistic relationships between each model variable and the target class better compared to other models, and supports the internal validity of the experiments further. Depending on how well an experiment or a tool measures the intended construct, a construct validity threat might be observed. To mitigate construct validity threats, the unconventional signal generation process is automated. External validity is related to the consistency of the claims made in a research study. It is hard to conclude that the results observed in this research could be generalized. Therefore, further research with additional unconventional signals and target entities is needed to justify our observations. 

\section{Conclusion}
\label{concl}
This paper uses unconventional signals extracted from the Twitter and GDELT data sources and shows that they can be used to predict various cyber attacks for the anonymized target entity KNOX. Furthermore, different time granularities (defined as $t_x$ and $t_g$) are used to calculate the unconventional signals and the entity ground truth for KNOX. Each time the signals are averaged over a period of time of length $t_x$, to predict the cyber incidents for the succeeding period of time of length $t_g$. Although the extracted signals are not explicitly related with the target entity KNOX, this paper shows that if appropriate $t_x$ and $t_g$ time granularities are chosen, the combined use of unconventional signals is able to achieve 70 \% AUC (Area Under ROC Curve) performance for Malware, Defacement, DOS, and Malicious Email/URL attack types for KNOX between April to November 2016. Although the set of signals that are related to the target attack class is different for each attack type, GEM, GET, and GEA signals seem to be more indicative of cyber incidents when compared to other signals used in this study. Moreover, this work also shows that using a different version of the same signal calculated with a different $t_x$ could improve the overall prediction model. 

Furthermore, common Machine Learning filtering methods like SMOTE and RandomSubsample are used to see if the attack prediction performance is improved when a skewed data set is balanced. This work developed SMOTE++ that uses a hybrid approach where under sampling, synthetic minority instance generation and reweighing techniques are used together to balance the distribution of the majority and minority instances in an imbalanced data set. We show that the performance of the Bayesian classifier on a skewed data set could be improved by making the data set balanced. We also show that the proposed SMOTE++ technique improves the prediction performance for most of the data sets for different attack types and the improvement amount is significant in some of these cases. As a future work we plan to use more signals from several other open data sources and include ground truth of different target entities. The proposed approach can be used with additional unconventional signals to achieve a higher forecast performance.

\section*{Acknowledgments}
This research is supported by the Office of the Director of National Intelligence (ODNI) and the Intelligence Advanced Research Projects Activity (IARPA) via the Air Force Research Laboratory (AFRL) contract number FA875016C0114. 

\bibliographystyle{ieeetran}
\bibliography{myreferences}

\vspace{-1 cm}
\begin{IEEEbiography}
    [{\includegraphics[width=1in,height=1.25in,clip,keepaspectratio]{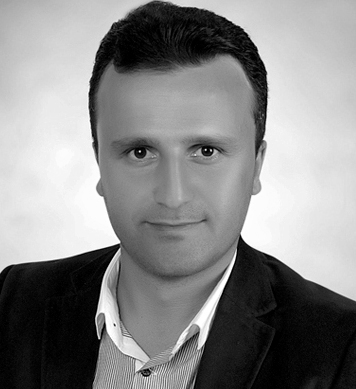}}]{Ahmet Okutan}
Dr. Ahmet Okutan received his B.S.degree in Computer Engineering from Bosphorus University, Istanbul, Turkey in 1998. He received his M.S. and Ph.D. degrees in Computer Engineering from Isik University, Istanbul, Turkey in 2002 and 2008 respectively. Dr. Okutan is currently a Post Doctoral Research Fellow in the Computer Engineering Department at Rochester Institute of Technology. He worked as analyst, architect, developer, and project manager in more than 20 large scale software projects. He has professional experience regarding software design and development, mobile application development, database management systems, and web technologies for more than 18 years. His current research interests include cyber attack forecasting, software quality prediction and software defectiveness prediction.
\end{IEEEbiography}
\vspace{-1 cm}
\begin{IEEEbiography}
    [{\includegraphics[width=1in,height=1.25in,clip,keepaspectratio]{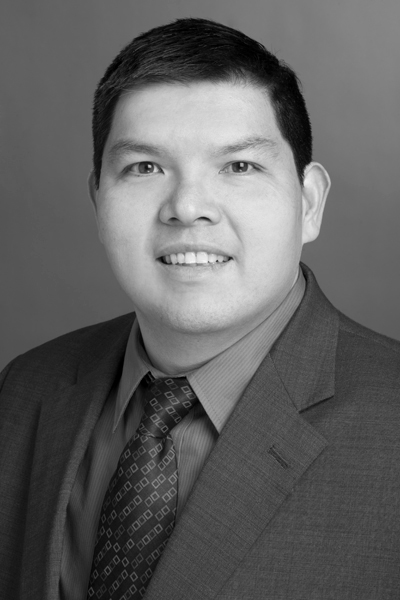}}]{Shanchieh Jay Yang}
Dr. S. Jay Yang received his B.S. degree in Electronics Engineering from National Chiao-Tung University, Hsin-Chu, Taiwan in 1995, and his M.S. and Ph.D. degrees in Electrical and Computer Engineering from the University of Texas at Austin in, 1998 and 2001, respectively. He is currently a Professor and the Department Head for the Department of Computer Engineering at RIT. He and his research group has developed several systems and frameworks in the area of cyber attack modeling for predictive situation, threat and impact assessment. He has published more than sixty papers and was invited as a keynote speaker, a panelist, and a guest speaker in various venues. He was a co-chair for IEEE Joint Communications and Aerospace Chapter in Rochester NY in 2005, when the chapter was recognized as an Outstanding Chapter of Region 1. He has also contributed to the development of two Ph.D. programs at RIT, and received Norman A. Miles Award for Academic Excellence in Teaching in 2007.
\end{IEEEbiography}
\vspace{-1 cm}
\begin{IEEEbiography}
    [{\includegraphics[width=1in,height=1.25in,clip,keepaspectratio]{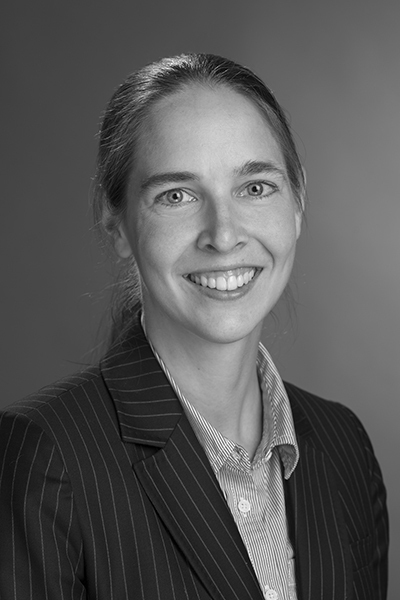}}]{Katie McConky}
Dr. Katie McConky received her B.S. and M.S. degrees in Industrial Engineering from Rochester Institute of Technology in 2005 and 2007 respectively, and her Ph.D. in Industrial Engineering from SUNY Buffalo in 2013. Dr. McConky is currently an Assistant Professor in the Industrial and Systems Engineering Department at Rochester Institute of Technology, and has seven years of industry experience working in the fields of information fusion and predictive data science prior to joining RIT. Currently Dr. McConky`s research work focuses on applying operations research and data analytic techniques to energy and security applications, including cyber security.
\end{IEEEbiography}

\end{document}